\documentclass[prb,twocolumn,amsmath,amssymb,showpacs]{revtex4}
\pdfoutput=1
\usepackage{graphicx}% Include figure files
\usepackage{dcolumn}% Align table columns on decimal point
\usepackage{bm}% bold math

\begin{document}

\title{Modeling a striped pseudogap state}
\author{Mats Granath$^1$ and Brian\ M.\ Andersen$^2$}
\affiliation{$^1$Department of Physics, University of Gothenburg,
SE-41296 Gothenburg, Sweden\\
$^2$Niels Bohr Institute, University of Copenhagen, DK-2100 Copenhagen, Denmark}

\date{\today}

\begin{abstract}

We study the electronic structure within a system of phase-decoupled one-dimensional superconductors coexisting with stripe spin and charge density wave order. 
This system has a nodal Fermi surface (Fermi arc) in the form of a hole pocket and an antinodal pseudogap. 
The spectral function in the antinodes is approximately particle-hole symmetric contrary to the gapped regions just outside the pocket. We find that states at the Fermi energy are extended whereas states near the pseudogap energy have localization lengths as short as the inter-stripe spacing. 
We consider pairing which has either local $d$-wave or $s$-wave symmetry and find similar results in both cases, consistent with the pseudogap being an effect of local pair correlations. 
We suggest that this state is a stripe ordered caricature of the pseudogap phase in underdoped cuprates with coexisting
spin-, charge-, and pair-density wave correlations. Lastly, we also model a superconducting state which 1) evolves smoothly from the pseudogap state, 2) has a signature subgap peak in the density of states, and 3) has the coherent pair density concentrated to the nodal region.

\end{abstract}


\maketitle

\section{Introduction}

The mysterious normal state pseudogap phase may hold the key to understanding the origin of high-temperature superconductivity in 
cuprate superconductors. The basic phenomenology of the pseudogap state is that of a partial suppression of the density of 
states (DOS) around the Fermi energy leaving only a residue of the expected Fermi surface.\cite{timusk,lee} The gap as mapped out by angle resolved
photoemission (ARPES) has an angular dependence which is similar to the $d_{x^2-y^2}$ form that is realized in the superconducting state with the exception of a remaining partial section of Fermi surface (so-called Fermi arc\cite{normanarc}) in the ``nodal'' part of the Brillouin zone (BZ).\cite{arpesreview} Therefore the pseudogap is ``antinodal'' in the sense that it has a maximum amplitude corresponding to the maximum amplitude of the superconducting gap.
Because of these similarites and because of the low superfluid density of these materials, it is natural to interpret the pseudogap in terms of phase disordered superconductivity.\cite{emerykivelson} This scenario agrees with a large Nernst signal detected at temperatures above T$_c$.\cite{ong} However, the fact that the pseudogap extends up to temperatures well above T$_c$ where no evidence of fluctuating superconductivity is found seems to rule out this scenario as an explanation of the overall pseudogap phenomenon. 
Another natural candidate for the opening of a spectral gap is generation of ordered phases, possibly in competition with the superconducting state. 

A different interpretation of the pseudogap is that it originates from a singlet-triplet gap, or spin gap, due to strong
nearest neighbor spin correlations related to the fact that the system is a doped Mott insulator.\cite{lee} This is the resonating valence bond (RVB) scenario.\cite{Anderson} Along these lines  
it has also been argued that stripes, which are unidirectional spin and charge
modulations, may play a key role and that the pseudogap would be a signature of a singlet pair correlation gap on the hole rich stripes.\cite{Emery,Zachar} Such a correlated state would be similar to the doped RVB state
but with the distinction of being preconditioned by the charge inhomogeneity. 

The existence of stripes is well established in La$_{2-x}$Ba$_x$CuO$_4$ (LBCO) and La$_{2-x}$Sr$_x$CuO$_4$ (LSCO) [with e.g. Nd co-doping] where both spin and charge Bragg peaks associated with the stripe order can be clearly identified, presumably enhanced by a structural transition in these materials.\cite{kivelsonRMP} More recently, however, evidence is accumulating that also other cuprates exhibit similar stripe correlations.\cite{hinkov,haug,xu} In this regard the universal hour-glass spin excitation spectrum  appears particularly striking,\cite{tranquadareview} and has a simple explanation within stripe models.\cite{stripe_neutron_theory}  Nevertheless, the importance and ubiquity of stripe correlations remain controversial, partly because of the complex disorder present in materials like Bi$_2$Sr$_2$CaCu$_2$O$_{8+\delta}$ (BSCCO) which is, however, very amenable to spectroscopic probes like ARPES and scanning tunneling microscopy/spectroscopy (STM/STS). 

It is well-known that ARPES finds a so-called momentum-space dichotomy consisting of coherent (incoherent) nodal (antinodal) quasiparticle states in the underdoped cuprates.\cite{arpesreview} Recent ARPES measurements have revealed the presence of two gaps in momentum space: 1) an antinodal (pseudo)gap which persists above $T_c$, and 2) a gap in the nodal region which exists only at $T<T_c$ and exhibits a temperature dependence consistent with a standard BCS gap.\cite{tanaka,kondo,wslee} More recently Yang {\it et al.}\cite{johnson}  found effectively two different types of gaps in 
the normal state at $T>T_c$, a particle-hole symmetric gap in the antinodal part of the BZ and a gap without particle-hole symmetry in the near nodal region. They also found evidence for a back bending dispersion in agreement with 
other recent work showing evidence for a nodal pocket instead of a disconnected arc.\cite{meng}
Consistent with a possible pairing origin of the antinodal gap, Kanigel {\it et al.}\cite{kanigel} identified the existence of particle-hole symmetric Bogoliubov bands in the pseudogap state.

STM experiments have consistently reported significant nano-scale gap modulations in the superconducting state with large-gap regions dominant in the underdoped regime where the pseudogap is known to be significant.\cite{cren,pan,howald,lang,gomes,boyer} Using the quasiparticle interference (QPI) technique, recent STM measurements on both BSCCO and Ca$_{2-x}$Na$_x$CuO$_2$Cl$_2$ (Na-CCOC) have revealed an electronic dichotomy similar to that found by ARPES: in the low-energy sector "conventional" dispersing quantum interference\cite{hoffman1,hoffman2} is detected from elastic scattering of coherent states near the nodal region, whereas incoherent states at energies near the antinodal gap give rise to a reduced set of non-dispersive peaks.\cite{howald2003,vershinen,hanaguri2004,mcelroy,hanaguri2007,liu,kohsaka2008,wise2008,jlee} In real-space the low-energy local density of states (LDOS) is largely homogeneous whereas higher bias real-space maps reveal characteristic heterogeneous checkerboard or locally stripe-like LDOS modulations. This data strongly suggests the presence of extended Bogoliubov states near the Fermi arc and quasi-localized states near the antinodal regions.\cite{mcelroycondmat,kohsaka2008} The spatially averaged DOS evolves smoothly from the superconducting state into the pseudogap phase above $T_c$,\cite{renner,boyer, pasupathy,jlee} where spatial nano-scale (pseudo)gap variations persist.\cite{vershinen,boyer} Finally recent pseudogap low-energy QPI STM studies of underdoped BSCCO\cite{jlee}  find a gapless Fermi arc and a particle-hole symmetric antinodal gapped region in qualitative agreement with the above mentioned ARPES experiments.\cite{johnson,kanigel}   

Theoretically, a stripe ordered pseudogap state was described in the extreme (and unrealistic) limit of decoupled one-dimensional correlated electron liquids.\cite{MG2001} Although such an approach can successfully classify the possible low temperature phases, it cannot readily incorporate a more realistic band structure in order to calculate a spectral function that can be compared to ARPES and tunneling experiments. However, within various approximate schemes, many calculations of the spectral distribution of a striped density wave have been 
performed.\cite{stripe_spectral_w,Granath2002,Baruch} Recently, these have attracted considerable interest in the context of high magnetic field transport measurements showing quantum oscillations possibly due to severe Fermi surface reconstruction from stripe ordering.\cite{Taillefer1,Taillefer2,normanmillis,kuchinskii,MG2008} This interpretation rely on electron-like single particle properties of bands which have predominant spectral weight in the antinodal region which seems at odds with the existence of an antinodal 
pseudogap.\cite{Tranquada,Sachdev}
It should also be noted that with realistic values of the Coulomb repulsion $U$, both standard mean-field theory and more sophisticated approaches  seem to indicate that there is no nodal spectral weight but only antinodal weight related to states centered on the hole rich stripes.\cite{Lorenzana} 

The more unsettling problem with these studies is the antinodal spectral weight itself. This is expected to be a robust feature, arising from the ``stripe bands'' which are the midgap states induced by the anti-phase domain walls in the antiferromagnetic spin texture.  For finite hole doping the spectral weight from these domain walls is antinodal for bond aligned stripes as we consider here, at least in mean-field. (For realistic stripe periodicity versus hole doping.) 
Thus, even though there is plenty of evidence that stripe correlations may be a generic feature of these materials it seems clear 
that such spin and charge density wave correlations will not give rise to an antinodal pseudogap.  

In this work we approach the pseudogap problem by formulating a non-correlated single-particle caricature of a striped pseudogap state by including a BCS pair term with finite amplitude only on the hole rich stripes and which is decoupled by phase disorder between stripes. In Hartree-Fock theory of the Hubbard model a striped state is realized by solving self-consistently for 
a collinear spin density wave (SDW) and a charge density wave (CDW) with the periodicity of the former being twice the latter.\cite{zaanen} 
It is in this framework that we will introduce a pairing term acting only in the high hole density regions of the CDW, which
is meant to mimic the actual singlet pair correlations below $T^*$ of a strongly correlated system. 

Our main result is that such a phenomenological model with stripe order and phase disordered on-stripe pairing is in fact broadly consistent with the salient spectroscopic features found by ARPES and STM. There is an approximately particle-hole symmetric antinodal spectral gap and a nodal region Fermi surface (Fermi arc) in the form of a hole pocket. 
In the near nodal region, just outside the pocket, the gap is due to SDW order and therefore not particle-hole symmetric. 
The DOS is suppressed in accordance with the antinodal gap and within an energy scale corresponding to the on-stripe pair potential. In this pseudogap state the gap is positioned at the Fermi level, and the DOS is roughly particle-hole symmetric but may have a minimum slightly above the Fermi energy due to the band dispersion giving the nodal pocket.The low-energy states that make up the Fermi surface are extended while states near the pseudogap energy are strongly localized on the order of the inter-stripe spacing. Finally, we also model a $d$-wave superconducting state coexisting with the spectral pseudogap and find a sub-gap peak in the DOS which is a result of gapping the nodal arc, and a coherent pair density that is concentrated to the nodal region.

\section{Model}
We consider a tight-binding model in a static field corresponding to unidirectional (striped) SDW order 
and a spatially modulated BCS pair field which is uniform along the stripe direction but not, in general, periodic in the direction transverse to the stripes. We will consider only bond centered charge period four stripes (spin period eight) and, in order to reduce the number of parameters, we do not include an explicit field that couples to the charge density.\cite{note_on_cdw} Here we will choose parameters such that there is a nodal hole pocket for realistic doping levels. In addition there will be antinodal weight in the form of open orbits or electron pockets.

On a square lattice with nearest neighbor hopping $t$, next-nearest neighbor hopping
$t'$, and chemical potential $\mu$ we have 

\begin{equation}
H_0=\sum_{\vec{k}\sigma}\xi_{\vec{k}}c^{\dagger}_{\vec{k},\sigma}c_{\vec{k},\sigma},
\end{equation}
with $\xi_{\vec{k}}=-2t(\cos(k_x)+\cos(k_y))-4t'\cos(k_x)\cos(k_y)-\mu$. 
We will take $t=1$ and use $t'=-0.3$. To generate a SDW we include the term 
\begin{eqnarray}
H_{SDW}&=& m\sum_{x,y,\sigma}\sigma(-1)^yV(x)n_{x,y,\sigma}\nonumber\\
&=&m\sum_{k_x,k_y,q,\sigma}\sigma V_q c^{\dagger}_{k_x,k_y,\sigma}c_{k_x-q,k_y-\pi,\sigma}\,.
\end{eqnarray} 
In the following, results are presented for bond-centered stripes with
$V(x\mod 8)=(.5,.5,-1,1,-.5,-.5,1,-1)$, giving finite $V_q$ for $q=\pm\pi/4$ and $q=\pm 3\pi/4$ only. The variation of the SDW is chosen such that the actual calculated spin density comes out with a similar variation. The spin potential will generate a CDW which has slightly higher hole density on the sites with aligned spin potential, which we will refer to as the stripes. 

\begin{figure}
\includegraphics[width=8.5cm]{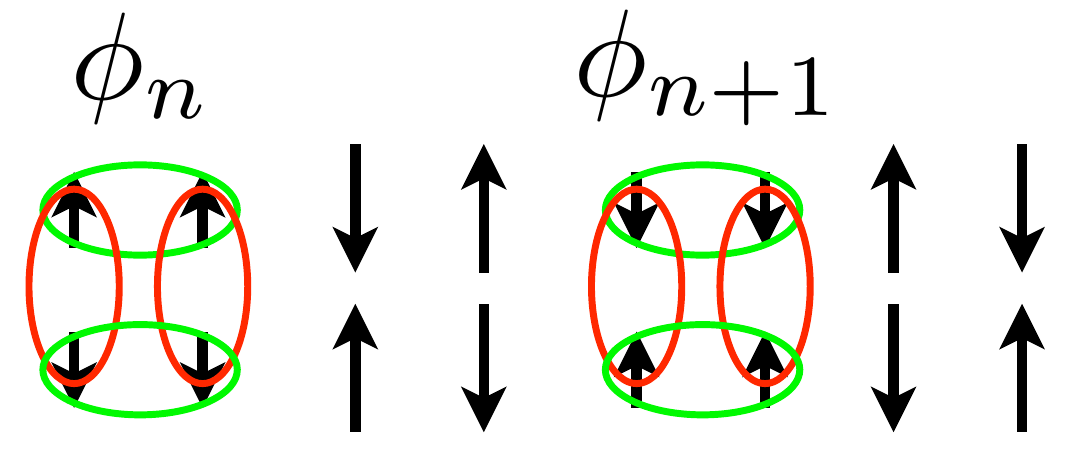}
\caption{\label{potentialfig}  (Color online) Unit cell of the SDW with the arrow length and direction proportional to the local spin together with the local $d$-wave paired bonds on two stripes with individual (and generally distinct) phases $\phi_n$ and $\phi_{n+1}$}
\end{figure}

The main purpose of the present study is to investigate the effect on the spectral distribution from local pairing that acts only on the stripes. For this purpose we include 
the following BCS term
\begin{eqnarray}
\label{HamiltonianBCSd}
H_{BCS,d}=\Delta_d \sum_{n,\langle y,y'\rangle,j=1,2} e^{i\phi_n}\Big[c^{\dagger}_{x_{j,n},y,\uparrow}c^{\dagger}_{x_{\overline{j},n},y,\downarrow}&-&\\
c^{\dagger}_{x_{j,n},y,\uparrow}c^{\dagger}_{x_{j,n},y',\downarrow}\Big]
&+&\mbox{H.c.}=\nonumber\\
2\Delta_d\!\!\!\! \sum_{k_x,k_y,q}\!\!\!\!  M_q(k_x,k_y)c^{\dagger}_{k_x,k_y,\uparrow}c^{\dagger}_{-k_x-q,-k_y,\downarrow}&+&\mbox{H.c.}\nonumber,
\end{eqnarray}
where $\langle yy'\rangle$ is sum over nearest neighbors along the stripe, $\phi_n$ is an arbitrary phase and $x_{j,n}$ with $j=1,2$ indicate the left ($j=1)$ and right ($j=2$) legs of the bond-centered stripe $n$, with $n=1,\ldots, N_x/4$ and $N_x$ the system size transverse to the stripe extension. The notation $\overline{j}$ refers to the leg opposite to $j$ on the stripe, i.e. $\overline{j}=1$ if $j=2$ and  $\overline{j}=2$ if $j=1$. Note that the pair creation and anihilation terms are symmetrized in real space coordinates and consequently spin singlet. (Nevertheless, as follows from symmetry considerations, in the presence of the SDW term there will in principle be induced triplet pair correlations at finite momenta. These are a natural consequence of the coexistence of SDW and pair correlations.\cite{Fenton})
In Eq.(\ref{HamiltonianBCSd}),
\[M_q(k_x,k_y)=e^{iq/2}D_x(q)\cos (k_x+q/2)-D_y(q)\cos (k_y)\] 
with $D_x(q)=(1/N_x)\sum_{n=1}^{N_x/4}e^{i(\phi_n-q 4n)}$ and 
$D_y(q)=(1/N_x)\sum_n(e^{i(\phi_n-q 4n)}+e^{i(\phi_n-q (4n+1))})$ is the Fourier transform of the local phases for the pairing along $x$ and $y$ directions respectively. We will mainly consider the problem with quenched uncorrelated random phases for which in general all
components $q$ of the pairing will be non-zero with a  periodicity $q\rightarrow q+\pi/2$. The above expression for $M_q$ is 
``$d$-wave like'' in the sense that the same term defined uniformly over the system would reduce to 
$M_q(k_x,k_y)=\delta_{q,0}(\cos(k_x)-\cos(k_y))$. Based on the underlying assumption of pair correlations tied to nearest neighbor singlet correlations and strong on-site repulsion we expect such a $d$-wave term to 
be the most realistic. Nevertheless, for comparison we also study ``$s$-wave like'' local pairing
\begin{eqnarray}
H_{BCS,s}=\Delta_s\!\!\!\! \sum_{n,y,j=1,2}\!\!\!\! ( e^{i\phi_n}c^{\dagger}_{x_{j,n},y,\uparrow}c^{\dagger}_{x_{j,n},y,\downarrow})+\mbox{H.c.}\nonumber\\
=\Delta_s\sum_{k_x,k_y,q} D_y(q) c^{\dagger}_{k_x,k_y,\uparrow}c^{\dagger}_{-k_x-q,-k_y,\downarrow}+\mbox{H.c.}\,,
\end{eqnarray}
with $D_y(q)$ as above. Figure \ref{potentialfig} shows the SDW field together with the $d$-wave paired bonds in a stripe unit cell.

The model $H=H_0+H_{SDW}+H_{BCS,\alpha}$ with $\alpha=d,s$ has long-range superconducting order along the stripe direction, which is clearly not realistic even in a stripe ordered system. Primarily this assumption is made to be able to numerically study large systems in the transverse direction
but may be partly motivated by the expectation of longer range pair correlations along the stripes. It constitutes an interesting project to extend the present model to include more realistic stripe disorder similar to the recent studies of Refs. \onlinecite{alvarez,robertson,delmaestro,kaul,andersen07}.

For system size $N_x$ the full Hamiltonian with a random phase $0\leq \phi_n<2\pi$ on each stripe $n=1,N_x/4$ corresponds to a $4N_x\times 4N_x$ matrix for each momentum $k_y$ (including $k_y+\pi$), giving a spectrum of Bogoliubov quasiparticles. 
From this we calculate 
the (zero temperature) spectral function $A(\vec{k},\omega)$ and the corresponding DOS 
$\rho(\omega)=\sum_{\vec{k}}A(\vec{k},\omega)$.  We average the spectral function over several independent realization of the random phases. 
We present results for magnitude of the SDW field $m=t/3$ and compare the cases without pairing, with $d$-wave pairing $\Delta_d=t/4$, and with $s$-wave pairing $\Delta_s=t/3$. The magnitude of $m$ is chosen such that for relevant doping levels, say $x<20$\%, the stripe ordered system exhibits nodal spectral weight. The values of the pairing are chosen to be in qualitative agreement with an experimentally measured pseudogap $\Delta_0$
which may be a large fraction of $t$. 

\section{Pseudogap state}
 
\begin{figure}
\includegraphics[width=8.5cm]{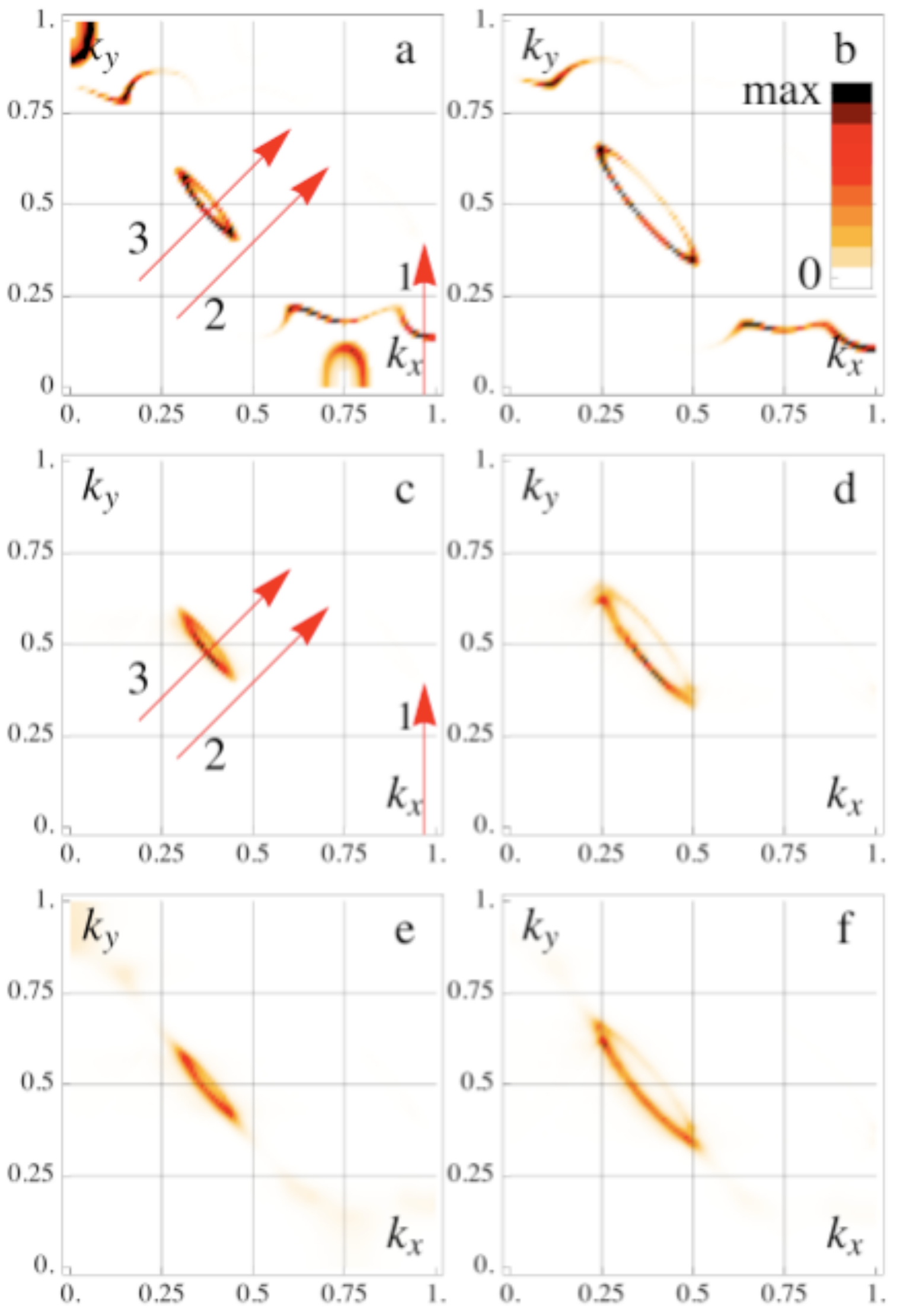}
\caption{\label{FSfig} (Color online) Intensity plot of the spectral weight at the Fermi energy in the first quadrant of the full BZ (units of $\pi/a$ with $a$ the lattice spacing) for $m=t/3$ and  $\mu=-1.0t$ (left column) or $\mu=-1.1t$ (right column).
 Top row (a, b) is without pairing, middle row (c, d) with phase disordered local $d$-wave pairing $\Delta_d=t/4$, and bottom row
(e, f) with phase disordered local $s$-wave pairing $\Delta_s=t/3$. Arrows in (a) and (c) indicate cuts shown in Fig. \ref{dispcuts}.
A Lorentzian broadening with width $\eta=0.01t$ was used. All plots have the same absolute color scheme. Calculations are made for a system size $N_x=200$ with periodic boundary conditions and averaged over three realizations of the quenched random phases on the charge stripes in (c-f). }
\end{figure}

\begin{figure}
\includegraphics[width=8.5cm]{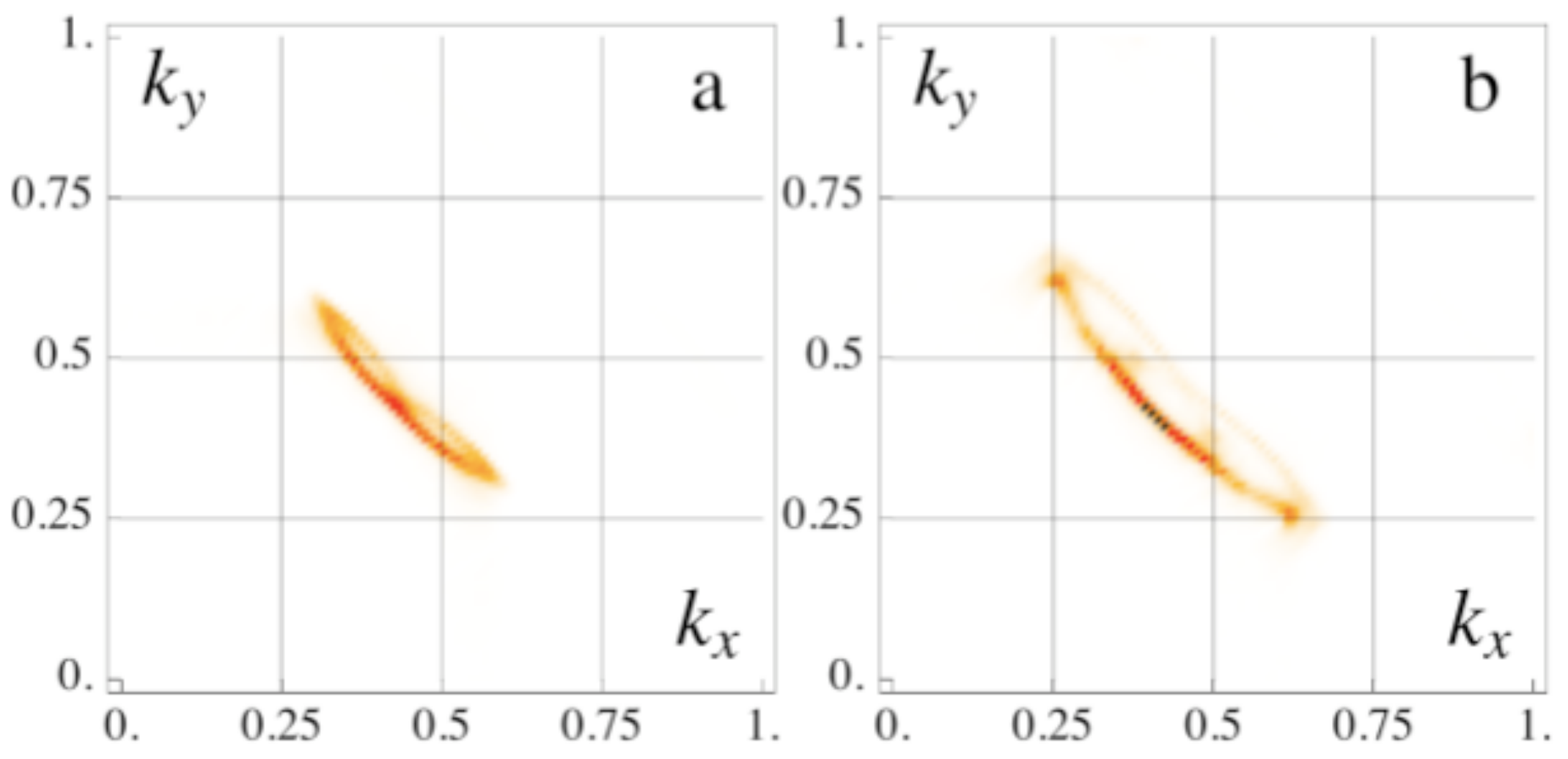}
\caption{\label{symmetrized_FS} (Color online) Spectral weight at the Fermi energy in the pseudogap state symmetrized with respect to the stripe orientation. The parameters are identical to Fig \ref{FSfig}c and \ref{FSfig}d, respectively. The ``arc'' length grows with doping but is in fact restricted by the Bragg planes at $k=\pi/4$ for the present case of period-eight spin stripes.}
\end{figure}

In the pseudogap phase, the low-energy spectral weight is presented in Fig. \ref{FSfig} for the three cases of 1) no pairing [Fig. \ref{FSfig}a,b], 2) local $d$-wave pairing [Fig. \ref{FSfig}c,d], and 3) local $s$-wave pairing [Fig. \ref{FSfig}e,f]. As seen from Fig. \ref{FSfig}a,b the stripe ordered system without any local pair correlations clearly exhibits antinodal spectral weight as found previously.\cite{stripe_spectral_w,Granath2002} However, the local pairing on the stripes removes this antinodal weight from the Fermi surface as seen from Fig. \ref{FSfig}c-f. This is a main result of this paper, local on-stripe 
pairing naturally gives an antinodal gap in the spectral function. At the same time the nodal hole pocket remains largely unaffected.
% with only very weak smearing from the phase-disordered pairing. 
Results for two different chemical potentials are presented in Fig. \ref{FSfig} [compare columns] showing that the nodal hole pocket grows with hole doping as expected. The actual hole density for these simulations come out to approximately 16\% and 20\% respectively with an average hole density variation of 2\% between hole rich (stripes) and hole poor regions (i.e. around 10\% variation around the mean). The spin density, $|\langle S^z_{x,y}\rangle |$ varies in magnitude between approximately 0.05 on stripes and 0.13-0.15 between stripes for these parameters. 

There is, at the level of the residual Fermi surface, no clear distinction between $d$- and $s$-wave pairing on stripes, indicating, as we will see, that it is the local nature of the pairing that is essential.
As discussed at the end of this section the phase disorder is essential in producing the gapless 
Fermi surface sections, a system with ordered phases will in general only have nodal points at the Fermi energy. 
In the rest of this section we analyze these results further.

As mentioned above, more realistic simulations of stripe correlations in the cuprate materials need to incorporate the role of quenched disorder and concomitant generation of short-ranged stripe domains.\cite{alvarez,robertson,delmaestro,kaul,andersen07} One important result of such disorder averaging is that the original fourfold rotational symmetry of the crystal lattice is restored globally. Within the present approach we can obtain a similar though admittedly more rough disorder-averaging by $x \leftrightarrow y$ symmetrizing our data. Performing such a symmetry operation of the spectral weight leads to the characteristic $k_x \leftrightarrow k_y$ symmetric Fermi arcs shown in Fig. \ref{symmetrized_FS}. As opposed to $(\pi,\pi)$ ordering scenarios such as conventional antiferromagnetic order or $d$-density wave order,\cite{ddw} within the present approach the Fermi arc is not symmetric around the AF zone boundary [lines connecting $(\pm \pi,0)$ and $(0,\pm \pi)$] in agreement with recent ARPES measurements in the pseudogap phase.\cite{johnson,meng} In principle, there is a back side of the hole pocket that might be detectable\cite{meng} but we anticipate that disorder will further reduce its weight.\cite{harrison,MG2008} 

\begin{figure}
\includegraphics[width=8.5cm]{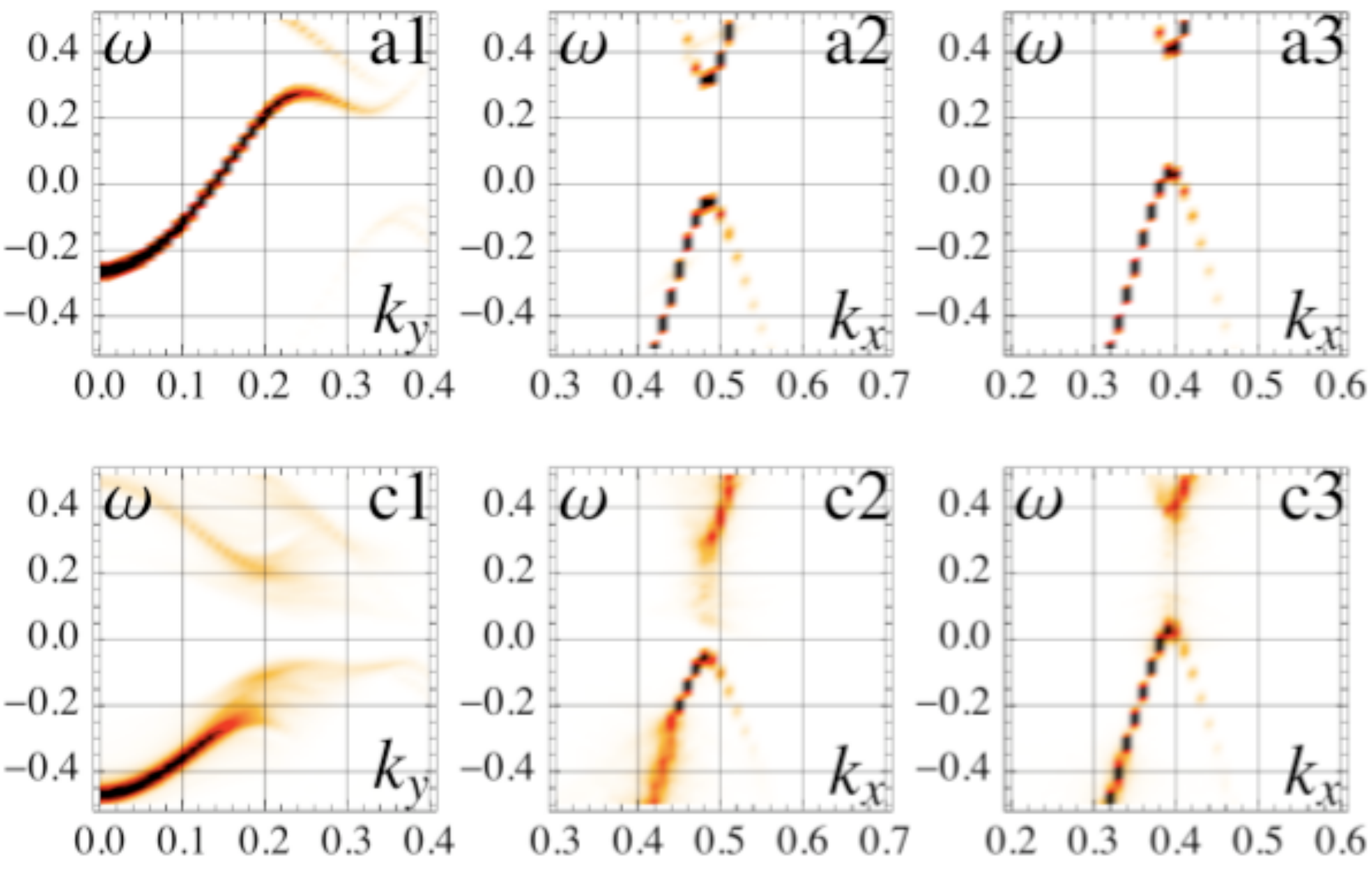}
\caption{\label{dispcuts} (Color online) Spectral weight along the three cuts indicated by arrows in Fig. \ref{FSfig}a and  \ref{FSfig}c. The top row (a1,a2,a3) [bottom row (c1,c2,c3)] corresponds to the parameters used in Fig. \ref{FSfig}a [\ref{FSfig}c]. The band in (a1) [along arrow 1 in Fig. \ref{FSfig}a] is split by the local pair potential $\Delta_d=0.25t$ giving a pseudogap of similar magnitude in the antinodal region (see also Fig. \ref{pseudogap_dos}). The cuts in the nodal region (a2,a3,c2,c3) [along arrows 2 and 3 in Fig. \ref{FSfig}a] are less affected by the pair potential and there is clearly (see e.g. c3) a band dispersing through the Fermi energy giving rise to the nodal hole pocket. 
Note the broadening as a signature of the pseudogap energy scale in near-nodal dispersion (c2). Energy $\omega$ in units of $t$ and with respect to the Fermi energy.}
\end{figure}

\begin{figure}[b]
\includegraphics[width=8.5cm,height=3cm]{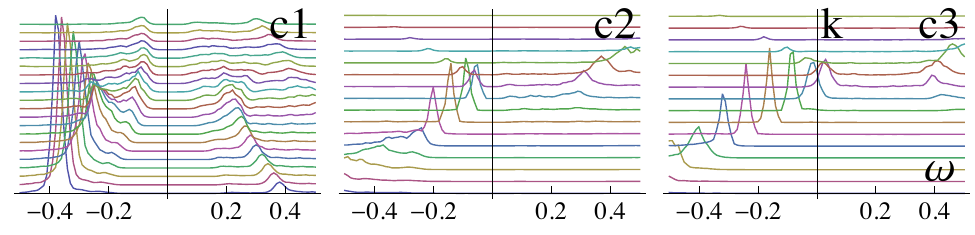}
\caption{\label{EDC} (Color online) Fixed momentum cuts of spectral function intensity from Fig. \ref{dispcuts}c1-c3.}
\end{figure}

\begin{figure}[t]
\includegraphics[width=8.5cm]{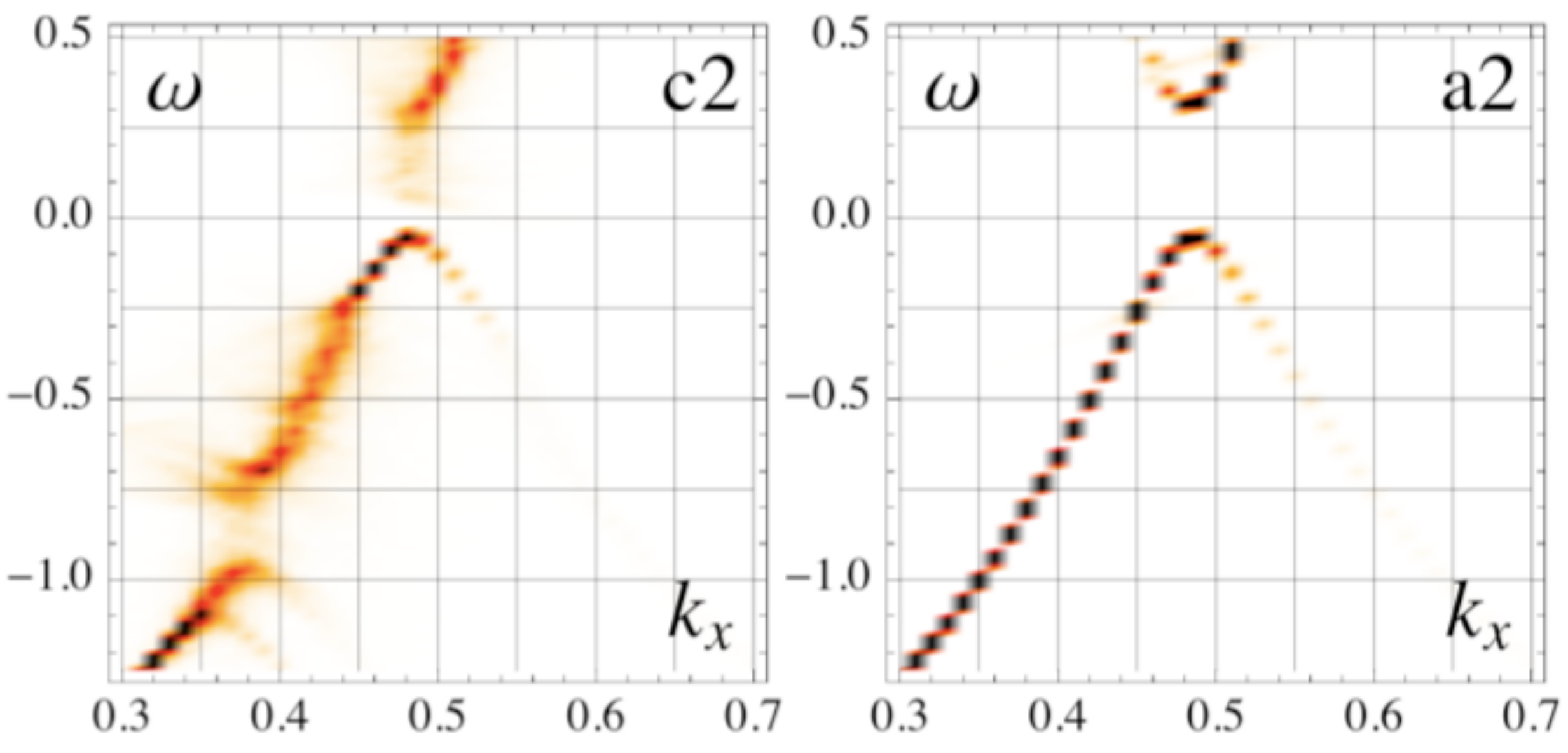}
\caption{\label{nodal_disp} (Color online) The same nodal dispersion as in Fig. \ref{dispcuts}a2 and \ref{dispcuts}c2 plotted over a larger energy range. One sees a small flattening of the low energy dispersion and a broadened high energy dispersion in c2 as compared to a2 with an abrupt change as a function of energy.}
\end{figure}

Figure \ref{dispcuts} displays several energy versus momentum cuts [see arrows in Fig. \ref{FSfig}a,c] of the spectral intensity further showing the effect of the disordered pair potential. For the antinodal cut in Fig. \ref{dispcuts}c1 a pairing gap is opened with a clear approximate particle-hole symmetry of finite spectral weight symmetrically around the Fermi energy. This is also evident from fixed momentum cuts (energy distribution curve (EDC)) in Fig. \ref{EDC}c1. Note that the minimum in the EDC curves in Fig. \ref{EDC}c1 remains at the Fermi level in the antinodal region in agreement with ARPES data and contrary to most other ordering scenarios for the pseudogap phase with non-zero ordering vector and no pairing correlations.\cite{normantheory} By contrast, the low energy nodal weight
is much less affected by the pairing as seen by comparing e.g. Fig. \ref{dispcuts}a3 to Fig. \ref{dispcuts}c3; in both cases there is a clear signature of a band dispersing through the Fermi energy giving rise to the nodal hole pocket. This is also seen from the near-nodal and nodal EDCs shown in Fig. \ref{EDC}c2 and \ref{EDC}c3, respectively. At energies below the pseudogap scale there is a signature of the disordered pairing in the form of a kink in the dispersion and a sudden severe broadening of the spectral function. This renormalization of the nodal dispersion in the pseudogap phase with local pairing is seen more clearly in Fig. \ref{nodal_disp}. Interestingly these features of the nodal dispersion are similar to what has been reported by ARPES measurements and discussed in the literature mainly in terms of bosonic mode-coupling scenarios.\cite{arpesreview}

\begin{figure}[b]
\includegraphics[width=8.5cm]{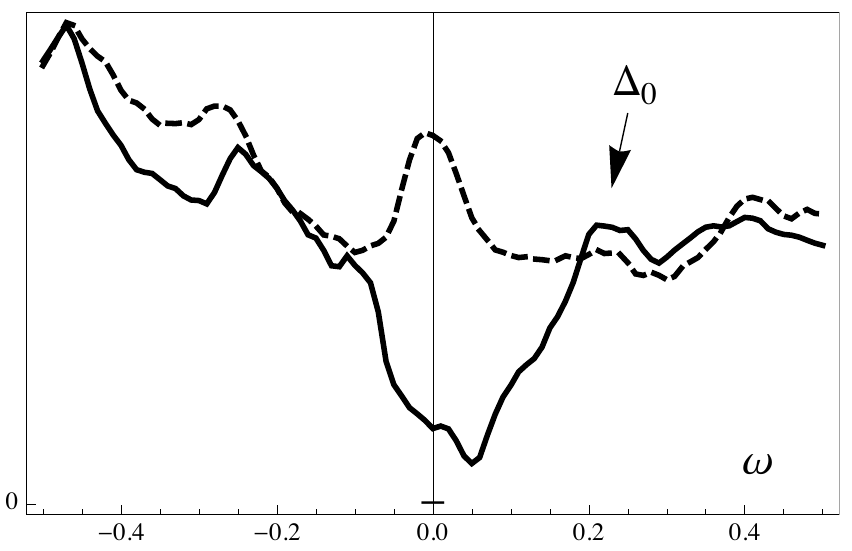}
\caption{\label{pseudogap_dos} Density of states without (dashed) and with (solid) local pair potential for the same parameters as in Fig. \ref{FSfig}a and \ref{FSfig}c, respectively. 
The magnitude of the local pair potential $\Delta_d=0.25 t$ is clearly reflected in suppression of the low-energy DOS below a pseudogap scale $\Delta_0\approx 0.25 t$ and is also seen 
to correspond to the splitting of the antinodal band (Fig. \ref{dispcuts}c1) around the Fermi energy. }
\end{figure}

\begin{figure}
\includegraphics[width=8.5cm]{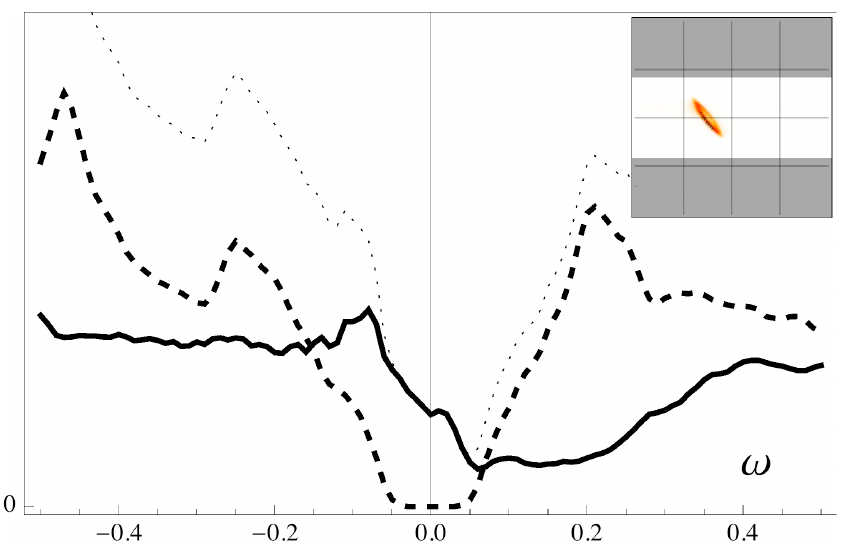}
\caption{\label{k_resolved_dos} (Color online) Nodal (solid), antinodal (dashed), and total (dotted) low-energy DOS. We find an approximately particle-hole symmetric antinodal pseudogap and a DOS consistent with a band dispersing through the Fermi energy in the nodal region. The nodal weight has a minimum at the edge of the nodal band where the pocket closes. Inset shows the spectral weight (c.f. Fig. \ref{FSfig}c) at the Fermi energy and the shaded (white) regions of the BZ are integrated over to get the antinodal (nodal) DOS.}
\end{figure}

In Fig. \ref{pseudogap_dos} we show the spatially averaged DOS in the case with (solid) and without (dashed) pairing correlations on the stripes. One can identify the pseudogap scale $\Delta_0\approx \Delta_d$ by the suppressed spectral weight below this energy. The magnitude of the pseudogap in the DOS
clearly corresponds to the main antinodal gap in the spectral function. Even though the overall shape of the calculated spatially averaged DOS in the pseudogap phase consists of a suppression of spectral weight at the Fermi level, an interesting feature is that the minimum is not necessarily tied to the Fermi energy but shifted slightly to positive bias $\omega>0$. This is a manifestation of the fact that the nodal band disperses through the Fermi energy and consequently will not have a minimum at $\omega=0$. This, we suggest, is a telltale sign of the quasi-particle nature and band dispersion of the nodal states. Interestingly, recent STM experiments may already have measured such a shift to positive bias in the pseudogap state.\cite{jlee,yazdaniprivat} In our simulations we can study nodal and antinodal contributions to the DOS directly by integrating the spectral weight over only nodal or antinodal regions. For this purpose we define
\begin{eqnarray}
\rho_{nodal}(\omega)&=&4\sum_{k_x=0}^{\pi}\sum_{k_y=k_o}^{\pi-k_0}A(\vec{k},\omega)\,,\\
\rho_{antinodal}(\omega)&=&\rho(\omega)-\rho_{nodal}(\omega)\,,
\end{eqnarray}
where $k_0$ is some appropriate division between nodal and antinodal parts of the BZ. Such a division of the low-energy spectral weight is natural based on the disconnected nature of 
the Fermi surface in the stripe ordered system as seen e.g. in Fig. \ref{FSfig}a and \ref{FSfig}c. 
Figure \ref{k_resolved_dos} displays this division of the DOS for the same system with phase disordered $d$-wave pairing as shown by the solid line in Fig. \ref{pseudogap_dos} (using $k_0=0.3\pi$). The nodal DOS is consistent with a band dispersing through the 
Fermi energy with the minimum 
($\omega\approx 0.05$) corresponding to closing of the hole pocket (see Fig. \ref{dispcuts}c3). By contrast, the antinodal DOS has the property of an approximately particle-hole symmetric pairing gap. 
From this division it is evident why the total DOS may have a minimum shifted to positive energies. 
As a point of principle this is an important example, however 
in general we do find that the DOS in the pseudogap phase for other parameter values may have a minimum at $\omega=0$. Even though there is a 
band dispersing through the Fermi energy it may have a suppressed DOS because it is affected by the pair potential. 

\begin{figure}
\includegraphics[width=8.5cm]{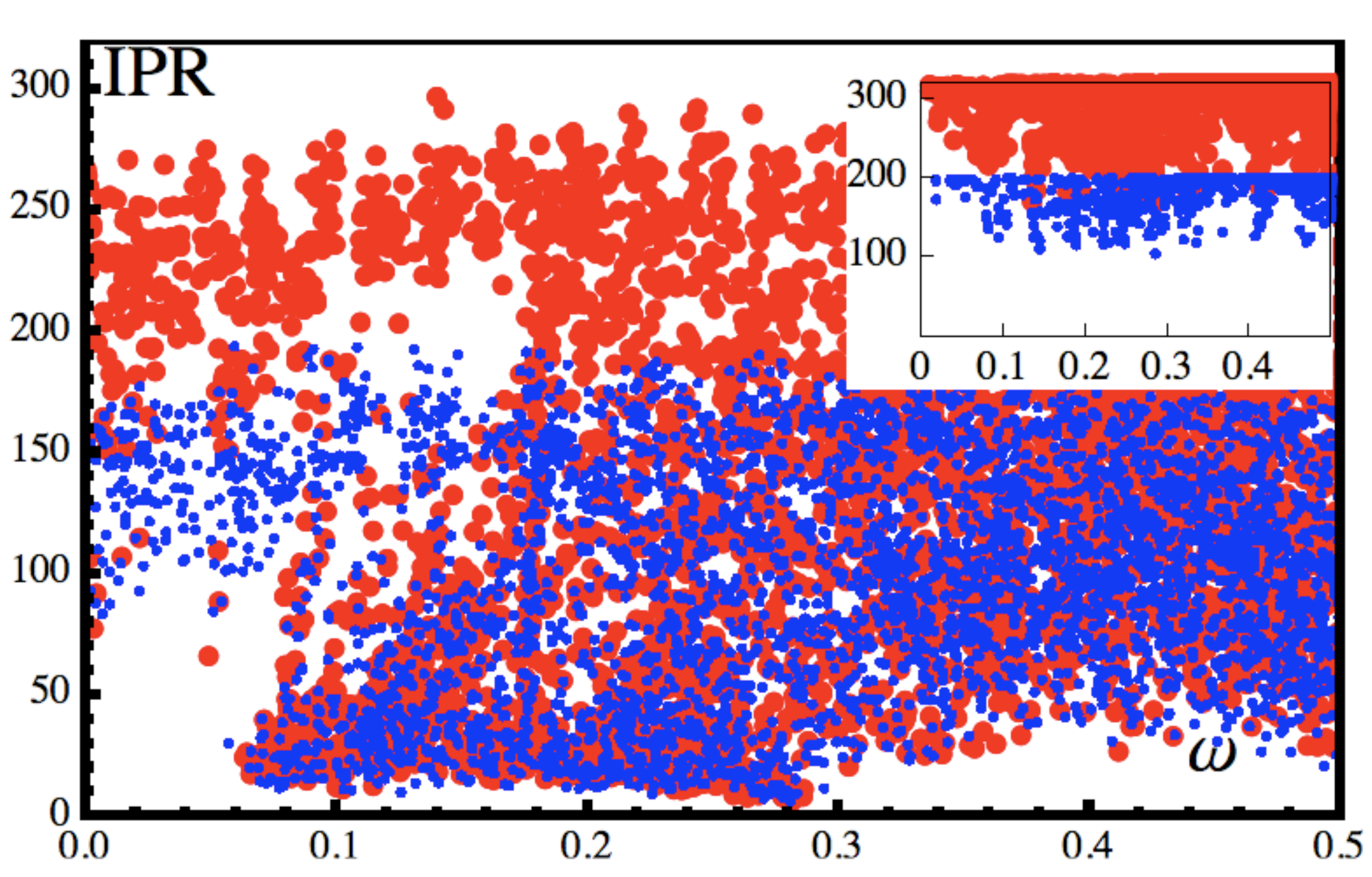}
\caption{\label{IPR} (Color online) One dimensional (direction transverse to stripes) IPR as defined in Eq. (\ref{IPR_eqn}) versus energy of quasiparticle excitations $E_\gamma$ for the same parameters as in Fig. \ref{FSfig}c with system size $N_x=200$ (small points) $N_x=320$ (large points). A large fraction of the states near the pseudogap energy $\Delta_0\sim\Delta_d=0.25 t$ are strongly localized in contrast to states near the Fermi energy. Inset is the corresponding IPR for the same parameters but with the same pairing phase on all stripes resulting in all states being extended Bloch states.}
\end{figure}

To further understand the nature of the low-energy states within this pseudogapped model system we study to what extent the quasiparticle excitations are localized as an effect of the disordered pair potential. To this end we calculate a one dimensional inverse participation ratio (IPR) since we can only have localization transverse to the stripe extension in the present model. Because of the local superconducting order, the quasiparticle excitations are Bogoliubons which give a contribution to the LDOS at both positive and negative energies. To get a measure of the extent of a state in the transverse ($x$) direction we add up these contributions and average along the stripe direction ($y$):
\begin{equation}
\mbox{IPR}_{\gamma}=(\sum_x[\sum_{y,\sigma}(\langle c_{x,y,\sigma}\gamma^{\dagger} \rangle)^2
+(\langle c^{\dagger}_{x,y,\sigma}\gamma^{\dagger}\rangle)^2]^2)^{-1}\,.
\label{IPR_eqn}
\end{equation}
Here the average $\langle \cdot \rangle$ is taken with respect to the ground state and $\gamma^{\dagger}$ is the creation operator of a quasiparticle excitation with energy
$E_\gamma$. A completely delocalized quasiparticle with equal weight on all sites (actually chains along $y$) will have $\mbox{IPR}=N_x$, with $N_x$ the system size, whereas a quasiparticle which only has spectral weight on a single site will have $\mbox{IPR}=1$. Figure \ref{IPR} is a plot of 
IPR as a function of quasiparticle energy of all states in an energy window $\omega<0.5 t$ for system sizes $N_x=200$ and $N_x=320$. As seen, states around the pseudogap energy $\Delta_d=0.25 t$ are most localized, some with an IPR as small as the stripe spacing and independent of system size, indicating that the states are basically localized on a single stripe (we expect localization length $\xi\sim$ $\mbox{IPR}/2$). Strikingly, the very lowest energy states (which make up the nodal hole pocket) all have a much larger IPR (here $>70$), and are clearly only weakly affected by the stripes and the disordered pair potential, in fact most of these states scale with system size. This is fully consistent with the band-like electronic quasiparticle nature of these states as discussed above. 

The transverse localization of the antinodal states at the pseudogap energy also makes it clear why the on-stripe pairing acts so strongly on these states; these states have most of their spectral weight on the stripes. This property of the antinodal stripe states was emphasized earlier for the case of diagonal disorder and related to the narrow band width in the transverse stripe direction and corresponding quasi-one dimensional nature of these states.\cite{MG2008,MG2006}
The role of the phase disorder in the present calculations is thus twofold, 1) it acts to localize antinodal states on stripes thus making these more susceptible to the local pairing, and 2) it causes the more extended nodal states which have most of the spectral weight between stripes to see only a very small spatially averaged pair potential, thus making these states relatively insensitive to the local pairing. %Alternatively, the ungapped nodal states can be thought of as a band of states that emerges from 
%a semi periodic set of localized states living on the domain walls of the phase shifted pair potential

\section{superconducting state}
\label{sc_section}

So far we have considered only phase disordered pairing, suggesting that this is a caricature of a correlated pseudogap state with local pair correlations in the form of an on-stripe spin gap. The most obvious extension of this model to a state with long-range superconducting order would be to lock the phases between stripes. Such a state would correspond to the mean-field theory of a stripe ordered superconductor with a periodically modulated superconducting order parameter, with finite components at $q=0$ and $q=\pm\pi/2$ (not $q=\pi$ for bond-centered stripes). We will not study this state here and only remark that in contrast to the phase disordered pairing, 
this state (for the $d$-wave case) has a Fermi surface consisting of points at the nodes of the gap function. For the particular case considered here (with pairing only on the bond centered stripes) we find a nodal line at $2\cos(k_y)-\cos(k_x)=0$ thus resulting in a point node that is shifted slightly away from that of an actual $d$-wave superconductor with intact fourfold rotational symmetry. 

Another alternative, similar to that suggested in the context of stripe ordered $x=1/8$ doped La$_{1.6-x}$Nd$_{0.4}$Sr$_{x}$CuO$_4$ La$_{2-x}$Ba$_{x}$CuO$_4$,\cite{himeda,berg} would be to lock the phases with a $\pi$ phase shift between neighboring stripes, resulting in a striped superconductor without a $q=0$ pairing component, and only finite momentum pairing. Because there is no zero momentum pairing such models have an extended Fermi surface similar to what we find here for the phase disordered pairing.\cite{Baruch}

In the following, we suggest a different model for the superconducting state which, in addition to the phase disordered pairing on stripes, also includes a uniform $d$-wave pair potential. The motivation for studying such a state comes from considering a glassy stripe state with static but only short range stripe order. In the superconducting state we expect there may be finite momentum pair correlations developing due to the local stripe order, but since phase coherence is established between stripe
patches with different orientations only the $q=0$ component may order. Thus in addition to the original model with  phase disordered on-stripe pairing we include explicitly the homogeneous component 
\begin{equation}
H_{BCS,h}\!=\!\Delta_h\!  \sum_{\vec{k}} [\cos (k_y)-\cos (k_x)] c^{\dagger}_{\vec{k}\uparrow}c^{\dagger}_{-\vec{k}\downarrow}+\mbox{H.c.}
\end{equation}

\begin{figure}
\includegraphics[width=8.5cm]{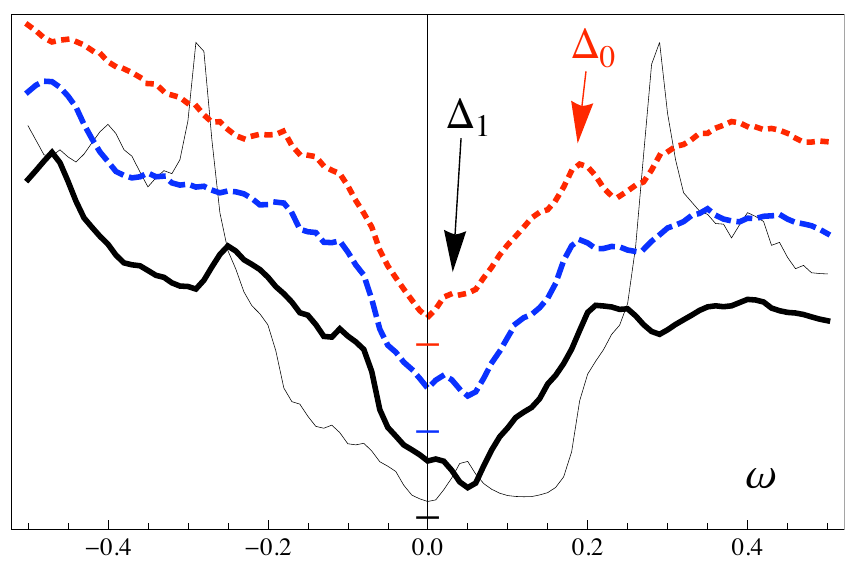}
\caption{\label{sc_dos} (Color online) Density of states (curves off-set for clarity) for different magnitudes of homogeneous $d$-wave pair potentials 
$\Delta_h=0$ (solid), $\Delta_h=0.05t$ (dashed), $\Delta_h=0.1t$ (dotted). The pseudogap energy is $\Delta_0$
and the energy $\Delta_1$ signifies the peak from the gapped nodal pocket in the superconducting state. The other parameters are the same as in Fig. \ref{FSfig}c: $m=t/3$, 
$\Delta_d=0.25t$, and $\mu=-t$. The thin line shows as comparison the DOS for a system
with only stripe order ($m=t/3$, $\Delta_d=0$, $\mu=-t$) and homogeneous $d$-wave potential $\Delta_h=0.15t$.}% (using $\eta=0.02$). 
\end{figure}

\begin{figure}[b]
\includegraphics[width=8.5cm]{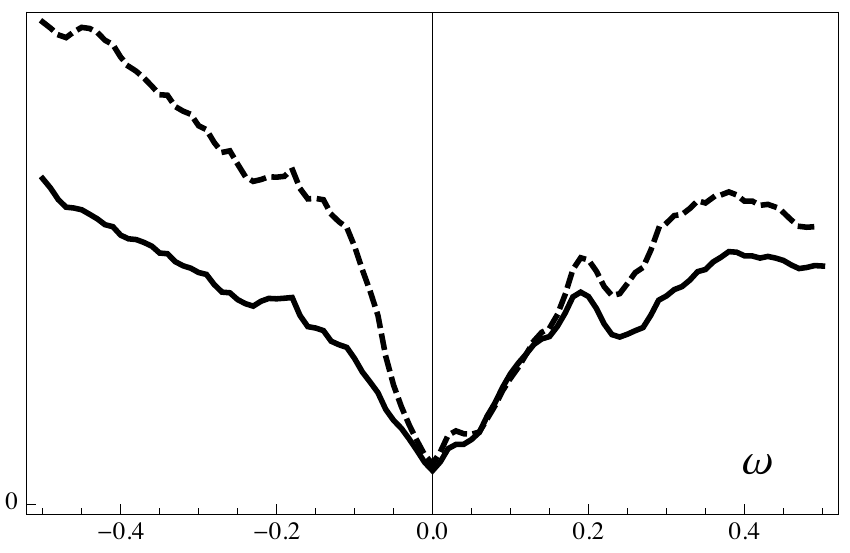}
\caption{\label{ldos} Local density of states in the superconducting state on (solid line) and off (dashed line) a charge stripe. The parameters used for this curve are similar to the dotted line in Fig. \ref{sc_dos}. The overall collapse of these two curves at low energy results in real-space homogeneity at low tunneling bias whereas the difference in LDOS at high energy will show up as stripes in the real-space STM field-of-view.}
\end{figure}

\begin{figure}
\includegraphics[width=8.5cm]{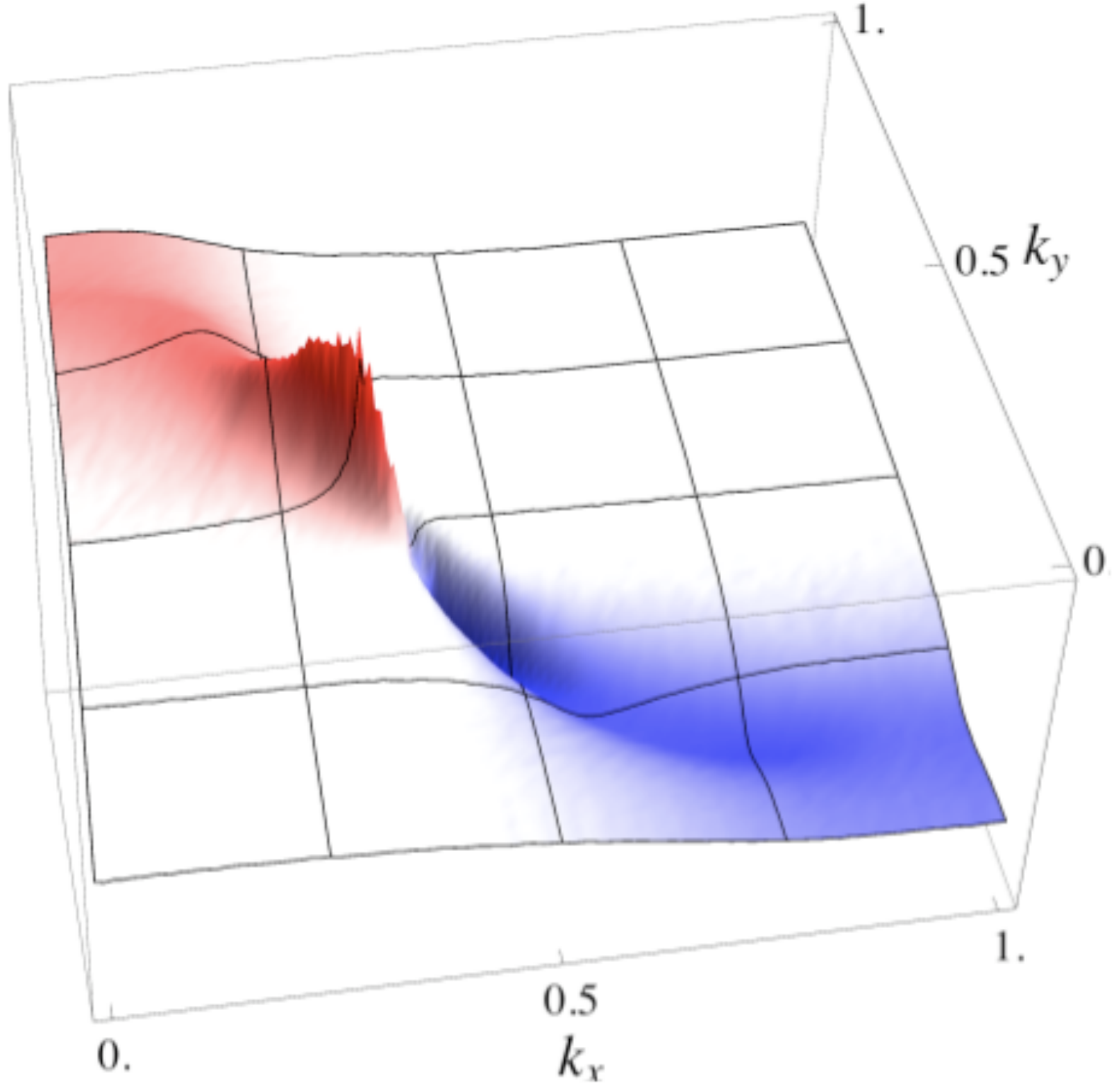}
\caption{\label{pair_density} (Color online) Zero mometum component of the pair density for parameters identical to the dashed curve in Fig. \ref{sc_dos}; $m=t/3$, $\Delta_h=0.05 t$ 
$\Delta_d=0.25t$, and $\mu=-t$. The shading indicates the deviation from zero. The weight is along the tight-binding Fermi surface with highest intensity on the nodal pocket of the pseudogap state.
The real part is plotted, there is also a smaller imaginary part (not shown) from the phase disordered on-stripe pairing.}
\end{figure}

This homogeneous pairing opens a $d$-wave gap on the Fermi arc. This is the actual gap of the superconducting condensate which will disappear in a standard mean-field fashion as temperature is increased.\cite{wslee}  
As a signature of the gapped pocket a sub-gap peak appears in the DOS as seen from Fig. \ref{sc_dos}. The origin and position of this peak is set by the transition of the contours of constant energy from separated ``bananas''  at low energy\cite{hoffman1,hoffman2} to a pocket at higher energy. This is similar to what has been discussed recently in the case of coexisting $d$-wave superconductivity and commensurate antiferromagnetism.\cite{andersen09} The energy of the subgap peak, $\Delta_1$, depends on band-structure and is not simply linearly related to the magnitude of $\Delta_h$ as seen in Fig. \ref{sc_dos} which shows the evolution of the DOS with increasing magnitude $\Delta_h$. The pseudogap energy scale, $\Delta_0$, that follows from the phase disordered pairing is maintained even in the presence of the homogeneous term as long as the latter is smaller in magnitude.  In contrast, a stripe ordered state with only a homogeneous pair potential has 
strong coherence peaks at twice the gap magnitude in addition to the smaller peak from the gapped nodal pocket as seen from the thin line in Fig. \ref{sc_dos}.

Figure \ref{ldos} shows the DOS spatially averaged over sites on (off) the charge stripes only, i.e the solid curve displays the DOS spatially averaged over sites that contain pseudogap bonds (see Fig. \ref{potentialfig}). One sees a low-energy "universal" LDOS consistent with homogeneous real-space STS spectra at low (sub-gap) bias, but deviations in the LDOS at higher energies resulting in spatially modulated (in this case stripy) LDOS real-space maps. The Fourier transform of the LDOS at a $q$ vector corresponding to the charge order, $q=2\pi/4$, is proportional to the difference between the curves shown in Fig. \ref{ldos}, and hence roughly particle-hole symmetric, even in the pseudogap state. 

Interestingly, a plot of the actual zero momentum pair density, $\langle c^{\dagger}_{\vec{k}\uparrow}c^{\dagger}_{-\vec{k}\downarrow}\rangle$, 
shown in Fig. \ref{pair_density} reveals that this quantity is dominated by weight in the nodal region on the hole pocket even though we consider a $d$-wave gap with larger magnitude in the antinodal regions. The reason for this is of course the coexisting antinodal pseudogap. This is consistent with a recent analysis of ARPES data by Kondo {\em et al.}\cite{kaminski} in which it was found that most of the coherent weight in the superconducting state is near the node. Similar conclusions about a coherent smaller gap below $T_c$ on top of the pseudogap have also been reached in earlier STM, ARPES and $\mu$SR work.\cite{boyer,tanaka,khasanov} Such a two gap phenomenology is naturally suggestive of a competition between a pseudogap state of non-pairing origin and a more traditional superconducting state. 
In the present model where the pseudogap is a pairing gap we have in mind an alternative scenario 
in which the pairing arises on the stripes which then infects the extended nodal states where the actual phase coherent pair density is formed. Although this mechanism is not contained in the model we may speculate that the antinodal spectral weight cannot couple coherently, possibly because the pseudogapped states are too localized or because the corresponding local pair density has most of the weight in finite momentum components.

\section{Discussion}

The model proposed in this paper does not fall into the conventional separation of scenarios for the pseudogap state: either preformed pairs or competing order. Our scenario contains both an order (SDW and CDW stripes) and concomitant local singlet pairing states which give rise to the spectroscopic antinodal pseudogap. Surprisingly, however, the Fermi arc on the nodal hole pocket is unaffected by this pseudogap mechanism irrespective of whether the local pairing symmetry is of $s$- or $d$-wave type. For the ordered SDW stripes studies here, the Fermi arc is the front side of the nodal hole pocket. However, in a more realistic short-range glassy phase the pocket is wiped out and replaced by a single arc.\cite{MG2008,harrison}  

As mentioned above, the phase disorder has negligible effects on the extended nodal states since the effective pairing potential averages to zero for these states. One can get additional understanding of the role of phase disorder by comparing to a stripe state with non-zero pairing and fixed phase [i.e. $\phi_n=0$ for all $n$ in Eq. (3)]. In such a phase-ordered stripe state, the pairing leads to nodal points at the Fermi energy, and from this respect the phase disorder is crucial for generating a Fermi arc. The origin of this increased spectral weight at the Fermi level can be traced to the fact that significant phase differences (e.g. $\pi$-phase shifts) between $d$-wave superconducting regions separated by antiferromagnetic regions generates low-energy states.\cite{andersen05,alvarezdagotto}

A recent study of thermal phase fluctuations of a $d$-wave superconductor also found the presence of a Fermi arc.\cite{altman} In that work too, the phase disorder is crucial for generating the arc, but contrary to the present approach, no pocket exists in the nodal region from SDW ordering. 
Even though the pocket could be destroyed by disorder one may be able to distinguish these scenarios for the Fermi arc by searching for particle-hole symmetry in the near nodal region, a property that should be absent in that region of momentum space within the approach presented here. (The latter in agreement with the recent analysis of ARPES measurements by Yang {\em et al.}\cite{johnson}) 

Experimentally it has also been found that the Fermi arc length scales with $T/T^*$.\cite{kanigelarclength} Even though extension of the present model to a self consistent study of finite $T$ effects is beyond the scope of this paper, we remark that a model of granular antiferromagnetic and $d$-wave superconducting islands found that the phase disorder can indeed generate a $T$-dependence of the Fermi arc,\cite{alvarezdagotto} similarly to the work by Altman {\em et al.}\cite{altman} For the present model it is natural for the arc length to vary inversely to the magnitude of the SDW potential
which in mean field is expected to grow with decreasing temperature. Whether or not such a variation can quantitatively 
reproduce the experimentally observed variations of arc length with temperature remains to be explored in more detail.\cite{MG_BMA_inprog}

Because the pseudogap proposed here remains at the Fermi level (modulo a possible small shift) it explains the smooth evolution of the spatially averaged DOS as a function of temperature from the superconducting state into the pseudogap state.\cite{renner,boyer,pasupathy,jlee} As we have shown, nodal states are extended whereas antinodal states are localized effectively on individual stripes in overall agreement with recent conclusions from e.g. tunneling spectroscopy.\cite{kohsaka2008} The nodal band can cause a slight shift of the DOS minimum to positive bias in the pseudogap state similar to what has been seen by recent STM measurements at $T>T_c$.\cite{jlee,yazdaniprivat} From the ARPES evidence of a quasiparticle band dispersion through the Fermi level together with an antinodal gap, this shift is something we expect on general grounds, and should be explored further by future tunneling experiments. 

The superconducting $q=0$ condensate at $T<T_c$ essentially "lives on top of" the pseudogap phase and its main spectroscopic effect is to gap the states at the Fermi arc. The superconducting gap gives rise to a sub-gap structure at $T<T_c$ within the pseudogap DOS. Interestingly, sub-gap features  in the spatially averaged DOS which disappear above $T_c$ have been recently pointed out by several STM experiments of underdoped cuprates.\cite{boyer,alldredge,pushp} A very interesting aspect of the present model for the superconducting state is that it naturally incorporates the existence of spatially varying local pairing amplitudes. This aspect was previously proven successful in describing salient features of the LDOS gap modulations measured by STM at $T<T_c$.\cite{mcelroy,mcelroyscience} Specifically, it reproduces the presence of sharp coherence peaks in small gap regions\cite{nunner1,nunner2} and the granular transition through $T_c$ observed by Gomes {\it et al.}\cite{gomes,andersen06} It is an interesting future study to extend  the present model to include more realistic disorder configurations and determine whether the spatially resolved LDOS is in further agreement with STM data. In the superconducting state we have seen that the low-energy LDOS is roughly constant (in space) within this approach but it is known that correlations which penalize charge fluctuations will further stabilize nodal LDOS universality.\cite{andersen08,garg} Also it will be interesting to see whether the present approach can reproduce the $k$-space dichotomy seen by Fourier transformed STM maps between the low- and high-energy momenta.\cite{jlee} In addition an essential aspect of the model is that the antinodal ``stripe'' states are localized by disorder in the phase of the local pair potential. An interesting question is whether similar localization may perhaps more realistically be caused by stripe density wave disorder.

Lastly, since the present model assumes SDW order it includes per construction local moments which, as shown recently, are present even in the magnetic response of the BSCCO materials.\cite{xu} The overall hour-glass neutron response is therefore fully compatible with the present model even though one may have to include glassy disorder and/or soft fluctuations to reproduce details of the measured low-energy magnetic fluctuations. 

\section{Conclusions}

We presented a phenomenological model for the pseudogap state consisting of stripe spin- and charge density wave order with phase decoupled on-stripe singlet pairing. This phase is characterized by extended states giving rise to a nodal hole pocket and gapped antinodal states that are localized transverse to the stripes. The spectral function exhibits particle-hole symmetry characteristic of a pairing gap in the antinodal region whereas in the near nodal region just outside the hole pocket the gap is due to stripe order and does not display particle-hole symmetry. The nodal dispersion has the appearance of a normal metal at the Fermi level but contains a kink and broadening below the pseudogap energy scale. The spatially averaged DOS has a gap at the Fermi level and is  also roughly particle-hole symmetric in the pseudogap state even though a small shift arising from the nodal band may be observable. Finally we also discussed a superconducting state coexisting with the pseudogap where the nodal pocket becomes gapped. Here we find a characteristic sub-gap peak in the DOS and a coherent pair density with most of the weight in the nodal region. 	  

\section{Acknowledgements}

The authors acknowledge useful discussions with J. C. Davis, P. J. Hirschfeld, E. W. Hudson, S. A. Kivelson, J. M. Tranquada, and A. Yazdani.
B. M. A. acknowledges support from The Danish Council for Independent Research $|$ Natural Sciences. M.G. thanks KITP-UCSB for hospitality under the program “The Physics of Higher Temperature Superconductivity” where parts of this work was completed. This work was supported in part by the National Science Foundation under the grant PHY05-51164 at the KITP.

\begin{figure}
\includegraphics[width=8.5cm]{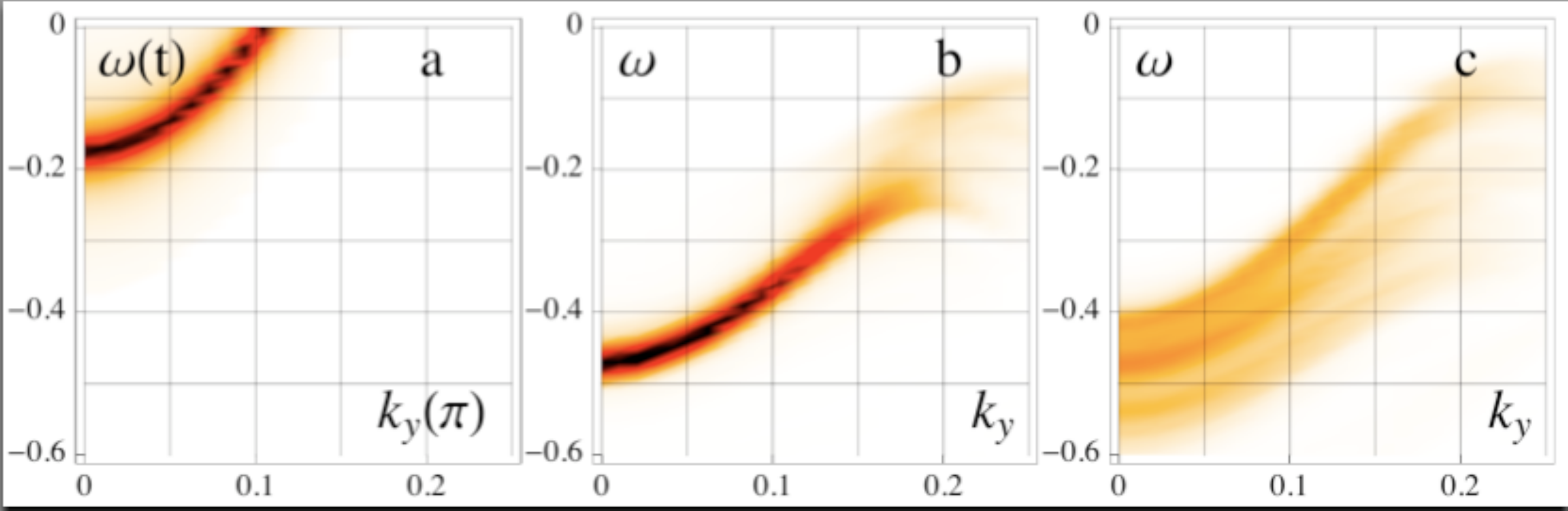}
\caption{\label{complementary} Antinodal spectral weight cut (1 in Fig.\ref{FSfig}), for a) bare band structure, $m=\Delta_d=\Delta_h=0$, b) pseudogap state $m=t/3$, $\Delta_d=t/4$, and c) superconducting state $m=t/3$, $\Delta_d=t/4$, $\Delta_h=0.1 t$ as discussed in the text. (b is identical to Fig. \ref{dispcuts} c1.)}
\end{figure}

{\em Note added.} After the submission of this work a new ARPES study of the pseudogap state in underdoped Bi2201 appeared.\cite{Hashimoto} 
An extended temperature regime ranging from above the pseudogap transition $T^*$ to below the superconducting transition was studied and found a conspicuous shift of the antinodal gap minimum momentum in the pseudogap state compared to $k_F$ above $T^*$ as well as an unexpected additional spectral broadening in the superconducting state. The observed shift suggests that the antinodal pseudogap is not (exclusively) a pairing gap. In light of this new ARPES data we realized that these qualitiative features were already present in the model considered in this paper. We thus present an additional Fig. (\ref{complementary}) using our earlier calculations showing a comparison between antinodal spectral weigth cuts for the bare band structure (modeling the region $T>T^*$), the pseudogap state, and superconducting state. There 
is a shift of the putative $k_F$ (turning point of coherent dispersion) in the pseudogap state compared to $k_F$ from the bare band structure, which is due to the stripe order. There is  also a broadening of the spectral weight in the superconducting state, which is related to the interplay between phase coherent zero momentum pairing and the local incoherent on-stripe pairing. The details of this broadening and the generics of the stripe induced shift of $k_F$ will be studied in more detail in subsequent work.  

\begin{thebibliography}{10}
\bibitem{note_on_cdw} The inclusion of a charge potential of similar or smaller magnitude than the spin potential is less essential because it does not introduce any additional Bragg planes and the spin potential anyway induces the corresponding charge modulation. An explicit charge potential will effect the various bands differently which may change the Fermi surface from containing hole pockets to electron pockets or both. See e.g. J. Lin and A. J. Millis, ArXiv:0908.2633. In this work we assume by construction a hole pocket and whether or not there are electron pockets or only open orbits in the antinodal region is not essential as these are gapped by the on-stripe pair potential.
\end{thebibliography}
\end{document}